\documentclass[epj]{svjour}
\usepackage{epsfig}
\begin{document}
\title{Hadrons in Hot and Dense Matter}
\author{Ralf Rapp\inst{1}
\thanks{\emph{Present address:} NORDITA, Blegdamsvej 17, DK-2100 Copenhagen;
work supported in part by US DOE grant no. DE-FG02-88ER40388.}%
}                     
%
%
\institute{Dept. of Physics and Astronomy, SUNY Stony Brook, 
New York 11794-3800, U.S.A.}
\date{Received: date / Revised version: date}
%
\abstract{The description of excitations in hot and dense 
(hadronic) matter is discussed with emphasis
on the use of correlation functions as a common framework
for comparing different model (and QCD lattice) calculations
with each other. Typical regimes of applicability of hadronic  
approaches are assessed, together with possibilities to confront 
them with experiment. We also elaborate on recent developments  
to relate baryonic in-medium effects to chiral symmetry restoration. 
} 
\maketitle
%
\section{Introduction}
\label{sec:intro}
The study of the phase structure of strong interactions ({\it i.e.},
QCD phase diagram) not only consists of determining the correct 
ground state, but also of identifying the relevant excitations, 
which are, after all, the probes amenable to experiment.  
At low to intermediate matter densities,
this corresponds to the widely investigated topic of
in-medium modifications of the hadronic spectrum as we know it
from the vacuum.
\begin{figure}[!h]
\begin{center}
\epsfig{file=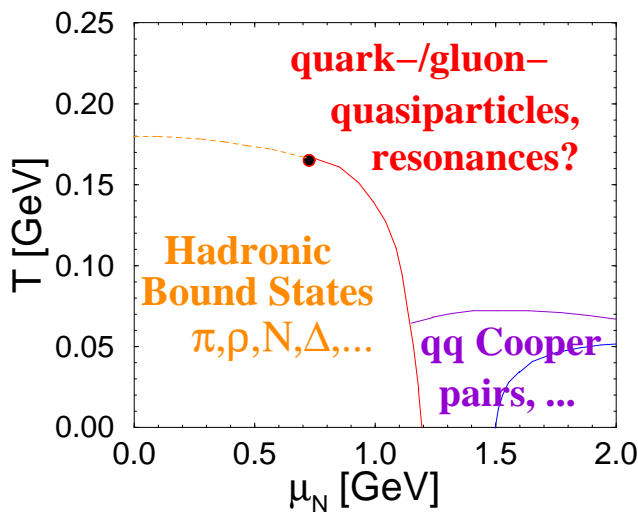,width=7cm,angle=-90}
\end{center}
\caption{Schematic phase structure and pertinent excitations 
of the QCD phase diagram.}
\label{fig:phase}
\end{figure}
When passing through a phase transition (or cross-over region)
an important question then is how changes in the ground state
symmetries (or order parameters) reflect themselves in the 
spectrum, schematically sketched in  Fig.~\ref{fig:phase}.
A general approach to the problem obviously requires a common 
framework which allows to compare results of individual models.

\section{Correlation and Spectral Functions}
\label{sec:corr}

\subsection{Definition}
The general description of an excitation in a given quan\-tum-number 
channel $\alpha$ can be formulated by correlation functions of 
the pertinent current operators~\cite{Shu93}, averaged over the 
thermodynamic partition function at temperature $T$ and (baryon-) 
chemical potential $\mu_B$.  In coordinate space one has  
\begin{eqnarray}
\Pi_\alpha(x)&=& {\rm tr} \left( [j_\alpha(x) j^\dagger_\alpha(0)]
 \ {\rm e}^{-(\hat{H}-\mu_B\hat{N})/T} \right) / {\mathcal Z} 
\nonumber\\
&=& \int \frac{d^4q}{(2\pi)^4} \ \Pi_\alpha(q) \ {\rm e}^{-iqx} \ . 
\end{eqnarray}
Its interpretation becomes more transparent in momentum space: 
in the timelike region, $M^2=q_0^2-\vec q^2 > 0$, the 
"spectral function" Im$\Pi_\alpha$ contains  
the information on the physical excitation spectrum, whereas the
spacelike region, $M^2<0$, encodes screening phenomena as well as 
fluctuation and susceptibility properties (see 
Refs.~\cite{CE01,PRWZ02} for recent examples in scalar and vector 
channel). 

In lattice QCD one typically evaluates $\Pi_\alpha(x)$ in
(Eucli\-dean) coordinate space; at finite temperatures,
the transformation into Minkowski space 
is complicated 
by the shrinking of the "Matsubara box" $\tau \epsilon \{0,1/T\}$ 
(due to periodic boundary conditions). 
However, model calculations in Minkowski space can, in principle,   
be straightforwardly transformed into Euclidean space for direct
comparison with lattice data, see Sect.~\ref{sec:vec-corr} 
below.

\subsection{Vacuum: Chiral Breaking vs. Confinement}
\begin{figure}[!h]
\begin{center}
\epsfig{file=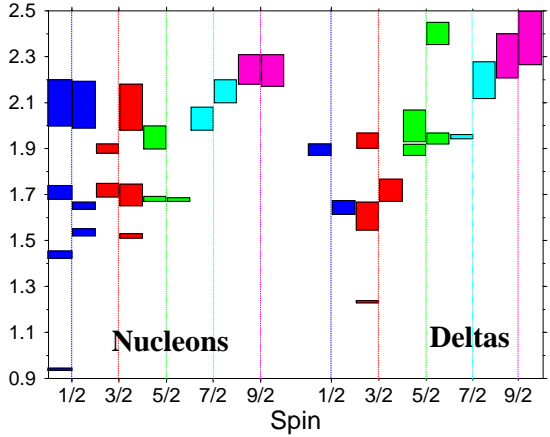,width=6cm,angle=-90}
\end{center}
\vspace{-0.3cm}
\caption{Baryon spectrum according to PDG~\protect\cite{pdg00}. 
For given spin, left and right column correspond to positive and negative 
parity, respectively. Note the 
mass gaps of $\Delta E\simeq 0.5$~GeV for low-lying 
spin-1/2 and -3/2 states, and the approximate degeneracy 
at higher masses, especially for $J\ge 5/2$.}
\label{fig:bar}      
\vspace{-0.5cm}
\end{figure}

Correlation functions in free space are determined 
by the propagation and interaction of 
anti-/quarks in the nonperturbative QCD vacuum, which is 
characterized by two main phenomena, confinement and spontaneous
breaking of chiral symmetry (SBCS). Whereas an understanding of the 
former is still elusive, the latter is widely believed to be due to 
instantons, which successfully reproduce QCD vacuum properties 
as well as low-lying meson and baryon correlation functions, see, 
{\it e.g.}, Ref.~\cite{SS96}. 
However, attempts to obtain confinement from instantons have
failed so far. On the other hand, the well-known Regge-behavior
of hadronic masses, $J\propto M^2$, is suggestive for  
string-like excitations associated with a confining force. 
Furthermore, at masses $M\ge 2$GeV, hadronic chiral partners seem 
to develop a degeneracy pattern suggestive for an "effective" 
restoration of chiral symmetry~\cite{CG01}.  
Most of these features are indeed supported by the appearance 
of the baryonic spectrum as extracted from the particle data 
group~\cite{pdg00}, cf.~Fig.~\ref{fig:bar}. 

\subsection{In Medium: Hadronic Approaches} 
Various methods to address medium effects on hadronic properties  
(and the equation of state) are available, cf.~Table~\ref{tab:1}.

At low densities and/or temperatures, systematic expansions using
chiral perturbation theory~\cite{KFW02} or a reduction
formalism~\cite{SYZ97} can be performed.

\begin{table}[!t]
\caption{Applicability domains of approaches to hot+dense
matter (ex = extrapolation, $n_0$=0.16~fm$^{-3}$).}
\label{tab:1}
\begin{tabular}{l|llll}
\hline \noalign{\smallskip}{\smallskip}
$T$ [MeV] & $\le$~120  & 120-150  & 150-$T_c$ & $\ge$~$T_c$ \\
\hline\noalign{\smallskip}{\smallskip}
$\varepsilon$ [GeV/fm$^{3}$] &  $\le$~0.05 & 0.05-0.3  & 0.3-0.8 & $\ge$~1 \\
\hline\noalign{\smallskip}{\smallskip}
$n_{had}$ [$n_0$] &  $\le$~0.5 & 0.5-2  & 2-5 & N/A \\
\hline\noalign{\smallskip}{\smallskip}
Method  & $\chi$PT & many-body & had-ex &  pQCD-ex  \\
\noalign{\smallskip}\hline\noalign{\smallskip}
\end{tabular}
\end{table}

At intermediate hadronic densities, as, {\it e.g.}, encountered in 
the later stages of heavy-ion collisions, 
many-body effects necessitating (partial) resummations, become  
important. {\it E.g.}, in the vector ($J^P=1^-$) channel, the 
"melting" of the $\rho$-meson~\cite{RW99} enables to describe 
the dilepton excess observed at CERN-SPS energies~\cite{ceres}
(cf.~Ref.~\cite{BR02} for an alternate view, and Ref.~\cite{RW00}
for a review).  
Likewise, in the scalar ($0^+$) channel, nuclear many-body effects 
contribute significantly to the threshold enhancement 
found in two-pion production experiments on nuclei~\cite{VO02}.   

In the expected phase transition region, both hadronic (below $T_c$) and 
partonic (above $T_c$) approaches have to rely on extrapolations. 
An important realization is, however, that medium modifications of 
light hadrons already encode precursor phenomena of chiral symmetry
restoration. This connection can be made explicit by addressing 
the modifications of the respective chiral partners on an equal 
footing ({\it e.g.}, $\pi$-$\sigma$, $\rho$-$a_1$, $N$-$N^*$(1535)). 

\section{Recent Applications: Axial-/Vector Channel}
\label{sec:appl}

\subsection{Dileptons and $a_1$ Spectral Function} 
\begin{figure}[!b]
\vspace{-2.2cm}
\epsfig{file=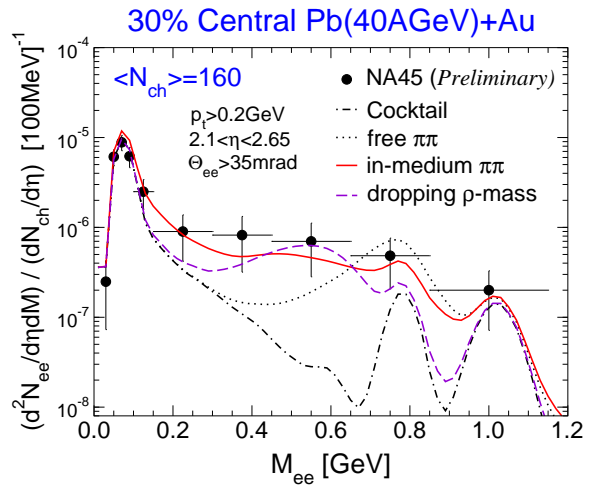,width=7cm}
\vspace{-0.3cm}
\caption{Theoretical predictions using in-medium $\rho$ 
mesons~\protect\cite{RW99} compared to recent NA45 
data~\protect\cite{ceres}.}
\label{fig:dl}
\end{figure}
There is general consensus that the large excess in low-mass 
($M$$\le$1GeV) dilepton production observed in central 
$Pb$(158~AGeV)+$Pb$ collisions by CERES/NA45~\cite{ceres} is 
due to in-me\-dium radiation from strongly modified $\rho$-mesons.
The importance of baryon-induced effects~\cite{RW99}
is supported by an even larger (relative) enhancement 
at lower SPS energies (40~AGeV), cf.~Fig.~\ref{fig:dl}. 
\begin{figure}[!th]
\begin{center}
\epsfig{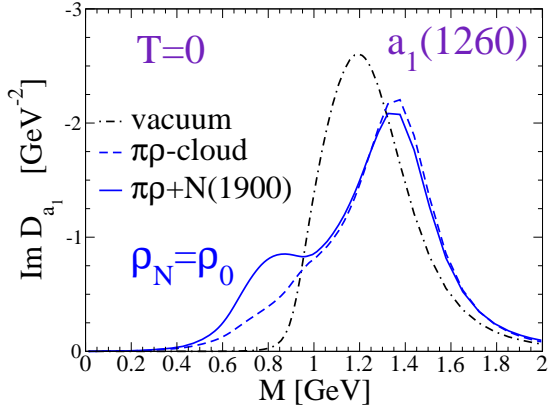}
\end{center}
\caption{$a_1$ spectral function in vacuum (dotted line)
and at normal nuclear matter density.}
\label{fig:a1}
\end{figure}
The final interpretation on the nature of chiral restoration
is still under debate. As argued above, this requires the evaluation
of the in-medium axialvector spectral function.
Such an analysis has been performed, {\it e.g.}, in Ref.~\cite{UBW02}
for a thermal $\pi\sigma\rho a_1$ gas. However, even in a {\em net}
baryon-free environment, the presence of baryon-antibaryon pairs 
at temperatures close to $T_c$ has appreciable impact on the 
$\rho$ spectral function~\cite{Ra01}. Preliminary results for 
the $a_1(1260)$ within a many-body framework at finite density 
are shown in Fig.~\ref{fig:a1}. 
In addition to an in-medium modified $\pi\rho$-cloud, significant
contributions arise from $a_1\to N^*(1900)N^{-1}$ excitations, 
the putative $SU(2)$ chiral partner of the $\rho\to N(1520)N^{-1}$ bubble.  
Experimental information on the decay modes of baryon resonances,  
here $N(1900)\to Na_1$ (possibly measurable with the HADES detector 
using $\pi$-beams at GSI), would be of great help to constrain
such calculations.

\subsection{Vector Correlator in Euclidean Space}
\label{sec:vec-corr}
We finally return to the question of model comparisons 
with lattice-QCD results. In a "mixed" representation, the 
temporal Euclidean correlator is related to the spectral function 
in Minkowski space via
\begin{equation}
\Pi_\alpha(\tau,q) = \int\limits_0^{\infty} 
\frac{dq_0}{\pi} \ {\rm Im} \Pi_\alpha(q_0,q)  
\frac{\cosh(q_0[\tau-\frac{1}{2T}])}{\sinh(q_0/2T)} \ . 
\end{equation}
A quenched lattice analysis of the vector correlator at zero
3-momentum, $\Pi_V(\tau,q=0)$, above $T_c$ has recently been 
presented in Ref.~\cite{Ka02}.
\begin{figure}[!ht]
\begin{center}
\epsfig{file=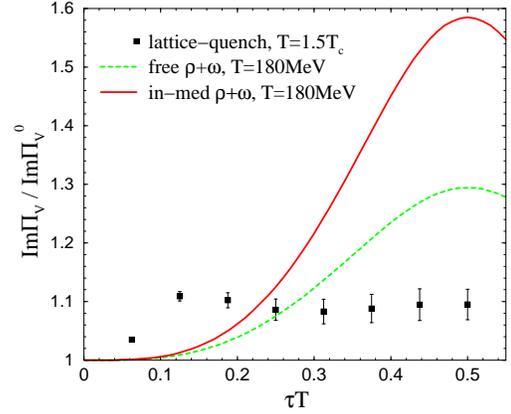,width=5.7cm,angle=-90}
\end{center}
\vspace{-0.2cm}
\caption{Temporal vector correlator at $T$=180~MeV(normalized to 
the ${\cal O}(\alpha_s^0)$ pQCD result) obtained from a hadronic 
model for $\rho$ and $\omega$ spectral functions~\protect\cite{Ra01}  
in vacuum (dotted line) and with medium effects at 
$\mu_B$$\simeq$25~MeV (solid line).   
Quenched lattice results (squares) at $T$=1.5$T_c$ are from  
Ref.~\protect\cite{Ka02}.}
\label{fig:vec}
\end{figure}
For illustration, we show in Fig.~\ref{fig:vec} results for
the same quantity using the sum of (free and in-medium) $\rho$ and
$\omega$ spectral functions, supplemented with a perturbative
continuum above $q_0\simeq 1.3$~GeV. We stress that the enhancement at 
large $\tau T\simeq$0.5 is at the origin of the dilepton excess 
at low-mass (cf.~solid line in Fig.~\ref{fig:dl}). 

\section{(Some) Open Issues}
The investigation of hadronic excitations and their medi\-um 
modifications provides important insights into nonperturbative 
properties of the QCD phase diagram.  In vacuum, confinement and 
chiral symmetry breaking establish themselves in different (mass-) 
regimes of the spectrum, whereas associated phase transitions seem 
to occur at the same critical temperature (whether this holds true 
at high baryon densities is an open question). 
In particular, the role of (baryonic) resonances requires further 
understanding: on the one hand, they importantly figure into medium 
effects on, {\it e.g.}, light (axial-) vector mesons (as inferred from 
dilepton spectra) and thus into (the approach towards) chiral 
restoration; on the other hand, their copious population close to 
$T_c$ significantly reduces the "jump" in the number of degrees of 
freedom in the transition to an (almost) ideal Quark-Gluon Plasma.    
Hadronic model calculations can thus be expected to remain an 
important tool to interpret both experimental and lattice data.


\end{document}